# A Pedagogical Evaluation and Discussion about the Lack of Cohesion in Method (LCOM) Metric Using Field Experiment.

Ezekiel Okike

School of Computer Studies, Kampala International University,
Kampala, Uganda 256, Uganda.

**Abstract**
Chidamber and Kemerer first defined a cohesion measure for object-oriented software – the Lack of Cohesion in Methods (LCOM) metric. This paper presents a pedagogic evaluation and discussion about the LCOM metric using field data from three industrial systems. System 1 has 34 classes, System 2 has 383 classes and System 3 has 1055 classes. The main objectives of the study were to determine if the LCOM metric was appropriate in the measurement of class cohesion and the determination of properly and improperly designed classes in the studied systems. Chidamber and Kemerer's suite of metric was used as metric tool. Descriptive statistics was used to analyze results. The result of the study showed that in System 1, 78.8% (26 classes) were cohesive; System 2 54% (207 classes) were cohesive; System 3 30% (317 classes) were cohesive. We suggest that the LCOM metric measures class cohesiveness and was appropriate in the determination of properly and improperly designed classes in the studied system.

## 1. Introduction

Software metric is any type of measurement that relates to a software system, process or related documentation. On the other hand, software measurement is concerned with deriving a numeric value for some attributes of a software product or process. By comparing these values to each other and to standards that apply across an organization, one may be able to draw conclusions about the quality of a software or software processes. The Lack of Cohesion in Methods (LCOM) metric was proposed in [5,6] as a measure of cohesion in the object oriented paradigm.

The term cohesion is defined as the "intramodular functional relatedness" in software [1]. This definition, considers the cohesion of each module in isolation: how tightly bound or related its internal elements are. Hence, cohesion as an attribute of software modules capture the degree of association of elements within a module, and the programming paradigm used determines what is an element and what is a module. In the object-oriented paradigm, for instance, a module is a class and hence cohesion refers to the relatedness among the methods of a class. Cohesion may be categorized ranging from the weakest form to the strongest form in the following order: coincidental, logical, temporal, procedural, communicational, sequential and functional.

i. Coincidental cohesion: A coincidentally cohesive module is one whose elements contribute to activities in a module, but with no meaningful relationship to one another. An example is to have unrelated statements bundled together in a module. Such a module would be hard to understand what it does and can not be reused in another program.

ii. Logical cohesion: A logically cohesive module is one whose elements contribute to activities of the same general category in which the activity or activities to be executed are selected from outside the module. A logically cohesive module does any of several different related things, hence, presenting a confusing interface since some parameters may be needed only sometimes.

iii. Temporal cohesion: A temporally cohesive module is one whose elements are involved in activities that are related in time. That is, the activities are carried out at a particular time. The elements occurring together in a temporally cohesive module do diverse things and execute at the same time.

iv. Procedural cohesion: A procedurally cohesive module is one whose elements are involved in different and possibly unrelated activities in which control flows from each activity to the next. Procedurally cohesive modules tend to be composed of pieces of functions that have little relationship to one another (except that they are carried out in a specific order at a certain time).

v. Communicational cohesion: A communicational cohesive module is one whose elements contribute to activities that use the same input or output data.

vi. Sequential cohesion: A sequentially cohesive module is one whose elements are involved in activities such that output data from one activity serve as input data to the next. Some authors identify this as informational cohesion.

vii. Functional cohesion: A functionally cohesive module contains elements that all contribute to the execution of one and only one problem-related task. The elements do exactly one thing or achieve one goal.

viii. A module exhibits one of these forms of cohesion depending on the skill of the designer. However, functional cohesion is generally accepted as the best form of cohesion in software design. Functional cohesion is the most desirable because it performs exactly one action or achieves a single goal. Such a module is highly reusable, relatively easy to understand (because you know what it does) and is maintainable. In this paper, the term "cohesion" refers to functional cohesion. Several measures of cohesion have been defined in both the procedural and object-oriented paradigms. Most of the cohesion measures defined in the object-oriented paradigm are inspired from the Lack of Cohesion in methods (LCOM) metric defined by Chidamber and Kemerer. In this paper, the Lack of Cohesion in methods (LCOM) metric is pedagogically evaluated and discussed with empirical data. The rest of the paper is organized as follows: Section 2 presents a summary of the approaches to measuring cohesion in procedural and object-oriented programs. Section 3 examines the Chidamber and Kemerer LCOM metric. Section 4 present the empirical study of LCOM with three Java based industrial software systems. Section 5 presents the result of the study upon which the LCOM metric was evaluated. Section 6 concludes the paper by suggesting that Chidamber and Kemerer's LCOM metric measures cohesiveness.

## 2. Measuring Cohesion in Procedural and Object oriented Programs

2.1    Measuring cohesion in procedural programs

Procedural programs are those with procedure and data declared independently. Examples of purely procedure oriented languages include C, Pascal, Ada83, Fortran and so on. In this case, the module is a procedure and an element is either a global value which is visible to all the modules or a local value which is visible only to the module where it is declared. As noted in [2], the approaches taken to measure cohesiveness of this kind of programs have generally tried to evaluate cohesion on a procedure by procedure basis, and the notational measure is one of "functional strength" of procedure, meaning the degree to which data and procedures contribute to performing the basic function. In other words the complexity is defined in the control flow. Among the best known measures of cohesion in the procedural paradigm are discussed in [3] and [4].

2.2 Measuring cohesion in object-oriented systems

In the Object Oriented languages, the complexity is defined in the relationship between the classes and their methods. Several measures exist for measuring cohesion in Object-Oriented systems [7,8,9,10,11,12]. Most of the existing cohesion measures in the object-oriented paradigm are inspired from the Lack of Cohesion in Methods (LCOM ) metric [5,6]. Some examples include LCOM3, Connectivity model, LCOM5, Tight Class Cohesion (TCC), and Low Class Cohesion (LCC), Degree of Cohesion in class based on direct relation between its public methods (DCD) and that based on indirect methods (DCI), Optimistic Class cohesion (OCC) and Pessimistic Class Cohesion (PCC).

## 3. The Lack of Cohesion in Methods (LCOM) Metric.

The LCOM metric is based on the number of disjoint sets of instance variables that are used by the method. Its definition is given as follows [5,6].

Definition 1.

Consider a class C1 with n methods $M_1, M_2,…,M_n$. Let $\{I_i\}$ = set of instance variables used by method $M_i$. There are n such sets $\{I_i\},…,\{I_n\}$. Let $P = \{ (I_i, I_j) | I_i \cap I_j = \phi\}$ and $Q = \{ (I_i, I_j) | I_i \cap I_j \neq \phi \}$. If all n sets $\{ I_1\}, …,\{I_n\}$ are $\phi$ then let $P = \phi$

LCOM = { |P|- |Q|, if |P| > |Q|

= 0, otherwise

Example: Consider a class C with three methods $M_1, M_2$ and $M_3$. Let $\{I_1\} = \{a,b,c,d,e\}$ and $\{I_2\} = \{a,b,e\}$ and $\{I_3\} = \{x,y,z\}$. $\{I_1\} \cap \{I_2\}$ is nonempty, but $\{I_1\} \cap \{I_3\}$ and $\{I_2\} \cap \{I_3\}$ are null sets. LCOM is (the number of null intersections – number of non empty intersections), which in this case is 1.

The theoretical basis of LCOM uses the notion of degree of similarity of methods. The degree of similarity of two methods $M_1$ and $M_2$ in class $C_1$ is given by:

$$\sigma(\ ) = \{I_1\} \cap \{I_2\}$$

where $\{I_1\}$ and $\{I_2\}$ are sets of instance variables used by $M_1$ and $M_2$. The LCOM is a count of the number of method pairs whose similarity is 0 (i.e, $\sigma(\ )$ is a null set) minus the count of method pairs whose similarity is not zero. The larger the number of similar methods, the more cohesive the class, which is consistent with the traditional notions of cohesion that measure the inter relatedness between portions of a program. If none of the methods of a class display any instance behaviour, i.e. do not use any instance variables, they have no similarity and the LCOM value for the class will be zero. The LCOM value provides a measure of the relative disparate nature of methods in the class. A smaller number of disjoint pairs (elements of set P) implies greater similarity of methods. LCOM is intimately tied to the instance variables and methods of a class, and therefore is a measure of the attributes of an object class.

In this definition, it is not stated whether inherited methods and attributes are included or not. Hence, a refinement is provided as follows [14]:

Definition 2.

Let $P = \phi$, if $AR(m) = \phi \ \forall \ m \in M_I(c)$

$= \{\{m_1,m_2\} \mid m_1,m_2 \in M_I(c) \land m_1 \neq m_2 \land AR(m_1) \cap AR(m_2) \cap A_I(c) = \phi \}$, else

Let $Q = \{\{m_1,m_2\} \mid m_1,m_2 \in M_I(c) \land m_1 \neq m_2 \land AR(m_1) \cap AR(m_2) \cap A_I(c) \neq \phi \}$

Then LCOM2( c) = $\{ |P| - |Q|$, if $|P| > |Q|$

$= 0$, otherwise

Where $M_I$ are methods in the class c and $A_I$ are the attributes (or instance variables) in the class c ; AR denote attribute reference

In this definition, only methods M implemented in class c are considered; and only references to attributes AR implemented in class c are counted.

The definition of LCOM2 has been widely discussed in the literature [6,9,11,14,16]. LCOM2 of many classes are set to be zero although different cohesions are expected.

3.1     Remarks

In general the Lack of Cohesion in Methods (LCOM) measures the dissimilarity of methods in a class by instancevariable or attributes. Chidamber and Kemerer's interpretation of the metric is that LCOM = 0 indicates acohesive class. However, for LCOM >0, it implies thatinstance variables belong to disjoint sets. Such a class maybe split into 2 or more classes to make it cohesive.

Consider the case of an n-sequentially linked methods as shown in figure 3.1 below where n methods are sequentially linked by shared instance variables.

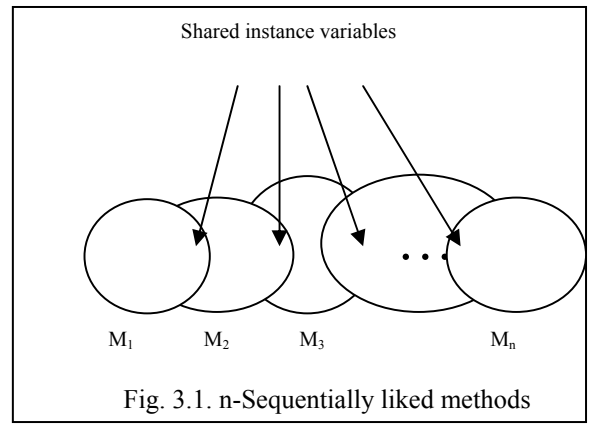

Fig. 3.1. n-Sequentially liked methods

In this special case of sequential cohesion:

$$P = \binom{n}{2} - (n-1) \quad (1)$$

$$Q = n - 1 \quad (2)$$

so that LCOM

$$P - Q = \left| \binom{n}{2} - 2(n-1) \right|_+ \quad (3)$$

where $[k]^+$ equals $k$, if $k>0$ and 0 otherwise [8].

From (1) and (2)

$$P - Q = \binom{n}{2} - (n-1) - (n-1)$$

$$= \binom{n}{2} - 2n - 2$$

$$= \binom{n}{2} - 2(n-1)$$

$$= \frac{n!}{(n-2)!2!} - 2(n-1) \qquad (4)$$

From (4), for $n < 5$, LCOM $= 0$ indicating that classes with less than 5 methods are equally cohesive. For $n \geq 5$, $1 < $ LCOM $< n$, suggesting that classes with 5 or more methods need to be split [8,18].

3.2 Class design and LCOM computation

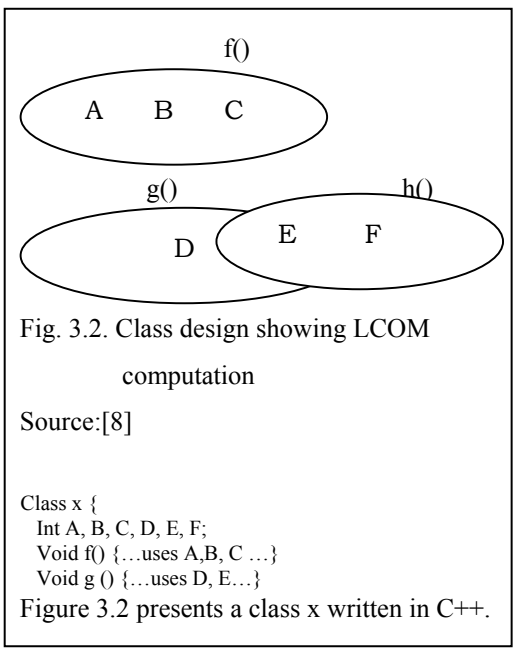

Fig. 3.2. Class design showing LCOM computation

Source:[8]

Class x {
  Int A, B, C, D, E, F;
  Void f() {…uses A,B, C …}
  Void g () {…uses D, E…}

Figure 3.2 presents a class x written in C++.

The Lack of Cohesion in Methods (LCOM) for class x = 1, calculated as follows:

There are two pairs of methods accessing *no* common instance variables namely (<f, g>, <f, h>). Hence P = 2. One pair of methods shares variable E, namely, <g, h>. Hence, Q = 1. Therefore, LCOM is 2 - 1 =1.

3.3 Critique of LCOM metric

The LCOM metric has been criticized for not satisfying all the desirable properties of cohesion measures. For instance, the LCOM metric values are not normalized values [11,13]. A method for normalizing the LCOM metric has been proposed in [18, 19]. It is also observed that the LCOM metric is not able to distinguish between the structural cohesiveness of two classes, in the way in which the methods share instance variables [8]. Hence, a connectivity metric to be used in conjunction with the LCOM metric was proposed. The value of the connectivity metric always lies between 0 and 1 [8].

## 4. The empirical study

### 4.1 The Method

Chidamber and Kemerer's suit of metrics namely: Lack of Cohesion in methods (LCOM), Coupling Between Object Classes (CBO), Response For a Class (RFC), Weighted Methods Per Class (WMC), Depth of Inheritance (DIT) and Number of Children (NOC) were used in the study. Two other metrics used in this experiment which are not part of the Chidamber and Kemerer metrics are: Number of Public Methods (NPM) and Afferent Coupling (CA). The choice of these metrics is informed by the need to have a metric to measure the number of public methods in a class as well as the number of other classes using a specific class. All the metrics used in this study provide the appropriate variables required for the experiments and the tools for measuring the metrics were readily available for use. In addition Chidamber and Kemerer's set of measure seems to be the basic set of object-oriented measures widely accepted [15].

Specifically cohesion was measured using the LCOM metric. Coupling was measured using CBO, RFC, and CA. Size was measured using WMC, and NPM. Inheritance was measured using DIT. Descriptive statistics was used to analyze results.

4.2     Description of variables

Table 4.1 below shows the variables used in the test systems. The metric of paramount interest is the LCOM although CBO, RFC, CA, WMC, NPM, NOC, and DIT values were obtained in order to verify if they are significant correlations between these and the LCOM.

Table 1: Metric variables used in the experiment

| Metric | Meaning | Attribute |
|--------|---------|-----------|
| LCOM | Lack of Cohesion in Methods | Cohesion |
| CBO | Coupling Between Objects | Coupling |
| RFC | Response For a Class | Coupling |
| CA | Afferent Coupling | Coupling |
| WMC | Weighted Method Per Class | Size |
| NPM | Number of Public Methods | Size |
| NOC | Number of Children | Inheritance |
| DIT | Depth of Inheritance | Inheritance |

• LCOM: Lack of Cohesion in Methods

Let P be the pairs of methods without shared instance variables, and Q be the pairs of methods with shared instance variables.

Then LCOM = |P| – |Q| , if |P| < |Q| i.e. If this difference is negative, LCOM is set to zero

• CBO. (Coupling between Object classes). A class is coupled to another, if methods of one class use methods or attributes of the other, or vice versa. CBO for a class is then defined as the number of other classes to which it is coupled. This includes inheritance based coupling.

• RFC (Response set for a class). The Response set for a class consists of the set M of methods of the class, and the set of methods directly or indirectly invoked by methods in M. In other words, the response set is the set of methods that can potentially be executed in response to a message received by an object of that class. RFC is the number of methods in the response set of the class.

• NPM (Number of Public Methods). The NPM metric counts all the methods in a class that are declared as public. It can be used to measure the size of an Application Program Interface (API) provided by a package [17].

• CA (Afferent Coupling). A class's Afferent Coupling is a measure of how many other classes use the specific class [17].

## 5.     Result and Discussion

The results of applying a Chidamber and Kemerer metric tool in the experimental study of the selected test systems consisting of 1472 Java classes from three different industrial systems are presented in this section. Descriptive statistics is used to analyze and interpret the results.

5.1     Descriptive statistics of the test systems

Descriptive statistics were used to obtain the minimum, maximum, mean, median, and standard deviation values for the test systems as shown in tables 2-4. In case of measurement for cohesion, the LCOM value lies between a range [0, maximum]. From Chidamber and Kemerer's interpretation of their LCOM metric, a class is cohesive if its LCOM =0. Using descriptive statistics, a median value in this range shows the level of cohesiveness in the system. This also means that at least half of the number of classes in the system are cohesive. The actual number of cohesive classes and their percentages based on the number of classes in the test systems were obtained from a simple frequency count of cohesive classes in each test system.

In this experiment, we applied a normalized LCOM ie [0,1]. This means that systems exhibiting high cohesion show low median values between [0,1]. From Chidamber and Kemerer's view, a median value of 0 indicates cohesive classes, however a median value of 1 is low enough to be a cohesive class. A minimum value indicates the lowest LCOM value for the class being measured. If this value is zero, it is the cohesion of the class from the LCOM interpretation. A maximum value indicates the highest LCOM value for the class. Using Chidamber and Kemerer's metric the LCOM values for a class can be any value from zero [0,1,2,3,..., 222,.. .6789..., maximum] to any value. The presence of such values as 222.. .6789... maximum make the LCOM metric not really appealing to most practitioners because a cohesion metric should not generate values which are not standardized (normalized). However Chidamber and Kemerer's position is that classes whose LCOM > 0 are improperly designed classes, and as such could be split to two or more classes to make them cohesive. The presence of outliers and un standardized (un normalized) values for LCOM is still a short coming with Chidamber and Kemerer LCOM metric. Using descriptive statistics, a maximum LCOM value indicates the value of the highest outlier in the measured system, and there could be more outliers within. Descriptive statistics for the test systems are shown in tables 2-4 . Table 5 provides the combine descriptive statistics for cohesion comparison across the test systems respectively.

Table 2: Descriptive statistics for system 1

| Statistics | WMC | DIT | NOC | CBO | RFC | 4d7 | CA | NPM |
|---|---|---|---|---|---|---|---|---|
| N | 34 | 34 | 34 | 34 | 34 | 34 | 34 | 34 |
| Valid Missing | 34 3 | 34 3 | 34 3 | 34 3 | 34 3 | 34 3 | 34 3 | 34 3 |
| Mean | 7.88 | 1.41 | .41 | 5.59 | 22.21 | 11 | 4.21 | 6.12 |
| Median | 4.00 | 1.00 | .00 | 5.00 | 18.50 | .00 | 2.00 | 4.00 |
| Stil. Dcv | 13.30 | .50 | 1.52 | 5.87 | 24.23 | 44 | 4.78 | 8.63 |
| Mm | 0 | 1 | 0 | 0 | 0 | 0 | 0 | 0 |
| Max | 74 | 2 | 8 | 31 | 129 | 2531 | 22 | 45 |

Table 3: Descriptive statistics system 2

| Statistics | WMC | DIT | NOC | CBO | RFC | LCOM | CA | NPM |
|---|---|---|---|---|---|---|---|---|
| N | 383 | 383 | 383 | 383 | 383 | 383 | 383 | 383 |
| Valid Missing | 383 0 | 383 0 | 383 0 | 383 0 | 383 0 | 383 0 | 383 0 | 383 0 |
| Mean | 8.24 | 2.14 | .58 | 8.33 | 20.93 | 150.40 | 5.70 | 6.97 |
| Median | 3.00 | 2.00 | .00 | 5.00 | 10.00 | 100 | 3.00 | 2.00 |
| Std Dev | 19.16 | 1.16 | 3.00 | 20.07 | 31.14 | 131.02 | 14.02 | 18.63 |
| Min | 0 | 1 | 0 | 0 | 0 | 0 | 0 | 0 |
| Max | 118 | 5 | 36 | 195 | 256 | 16290 | 157 | 181 |

Table 4: Descriptive statistics system 3

| Statistics | WMC | DIT | NOC | CBO | RFC | LCOM | CA | NPM |
|---|---|---|---|---|---|---|---|---|
| N | 1055 | 1055 | 1055 | 1055 | 1055 | 1055 | 1055 | 1055 |
| Valid | 1055 | 1055 | 1055 | 1055 | 1055 | 1055 | 1055 | 1055 |
| Missing | 0 | 0 | 0 | 0 | 0 | 0 | 0 | 0 |
| Mean | 7.96 | 1.42 | .36 | 6.25 | 26.49 | 44.91 | 1.69 | 5.91 |
| Median | 5.00 | 1.00 | .00 | 3.00 | 16.00 | 6.00 | .oo | 4.00 |
| Std.Dev | 9.40 | .62 | 3.00 | 7.55 | 30.93 | 180.45 | 5.83 | 6.82 |
| Min | 0 | 1 | 0 | 0 | 0 | 0 | 0 | 0 |
| Max | 109 | 4 | 64 | 65 | 210 | 2744 | 71 | 61 |

### 5.2 Cohesion comparisons across systems.

Table 5 below shows the comparison of cohesion measures across the three test systems. The actual number of cohesive and uncohesive classes per system and their percentages are indicated as shown above. A median value in range [0,1) indicates the system is cohesive.

Table 5: Cohesion comparison across the test systems

| SYSTEMS | NO. OF CLASSES | DESCRIPTIVE STATISTICS | | |
|---|---|---|---|---|
| | | LCOM | COHESIVE | UNCOHESIVE |
| System1 | 34 | Minimum 0<br>Maxim 2534<br>Mean 79.03<br>Median 0<br>Std.dev 440.22 | 26<br>(78.8%) | 7<br>(21.2%) |
| System2 | 383 | Minimum 0<br>Maxim 16290<br>Mean 150.40<br>Median 1.00<br>Std.dev 1318.35 | 207<br>(54%) | 176<br>(46%) |
| System3 | 1055 | Minimum 0<br>Maxim 2744<br>Mean 44.91<br>Median 6.0<br>Std.dev 180.45 | 317<br>(30%) | 738<br>(7%) |
| TOTAL | 1472 | | | |

### 5.3 Discussion

Following Chidamber and Kemerer's guide to interpreting their LCOM metric using descriptive statistics (minimum, maximum, mean, median, standard deviation) [6], a low median value indicates that at least 50% of the class have cohesive methods. In their original work this low median value was 0. However, a median value of 1 (also low) was considered in this work. In the field experiments result tables 2 — 4, it was observed that LCOM median values for systems 1 and 2 are 0.00 or 1.00. Hence these systems are considered to have more cohesive classes than system 3 whose LCOM median value is 6.00. To confirm this, a simple frequency count of cohesive and un-cohesive classes was carried out to find the actual percentages as shown in table 5. However, Chidamber and Kemerer's view that a class is not cohesive when LCOM= 1 does not seem appropriate as there was no reason to suggest that classes with LCOM =1 are improperly designed.

Since the LCOM metric is an inverse cohesion measure: a low value indicates high cohesion and vice versa [14]. For illustration, suppose the cohesion of a class ci is 0 (LCOM $(c_1)$ 0), and the cohesion of another class c2 is 1 (LCOM $(c_2)$= 1); this should mean that LCOM $(c_i)$> LCOM $(c_2)$, and should not be interpreted to mean that LCOM $(c_2)$=1 is not cohesive and therefore may be split.

### 6. Conclusion

In this paper the concept of cohesion in both the procedural and object-oriented paradigm has been extensively discussed. It is suggested that Chidamber and Kemerer's Lack of Cohesion in Methods (LCOM) metric measures cohesiveness. However, the presence of outliers and not standardized values make the metric not as appealing as its variant measures whose cohesion descriptive statistics values are standardized (normalized).

A normalized LCOM metric is already proposed in [19]. The metric may be used to predict improperly designed classes especially when the LCOM metric is used with reference to the Number of Public Methods (NPM) being greater than or equal five (NPM  5) [18,5,6]. Cohesion as an attribute of software when properly measured serves as guiding principle in the design of good software which is easy to maintain and whose components are reusable [3,4,5,18]

**Ezekiel U. Okike** received the BSc degree in computer science from the University of Ibadan Nigeria in 1992, the Master of Information Science (MInfSc) in 1995 and PhD in computer science in 2007 all from the same University. He has been a lecturer in the Department of Computer Science, University of Ibadan since 1999 to date. Since September, 2008 to date, he has been on leave as a senior lecturer and Dean of the School of Computer Studies, Kampala International University, Uganda. His current research interests are in the areas of software engineering, software metrics, compilers and programming languages. He is a member of IEEE computer and communication societies.